\documentstyle[12pt,epsfig]{article}
\renewcommand{\bar}[1]{\overline{#1}}

\begin{document}
\bigskip\bigskip
\begin{center} 
{\large \bf T-odd Observables in Elastic Scattering,
\\
in Deep Inelastic Processes and in Weak Decays}
\end{center}
\vspace{8pt}
\begin{center}
\begin{large}

E.~Di Salvo$^{1,2}$\footnote{Elvio.Disalvo@ge.infn.it},
Z.~J.~Ajaltouni$^{1}$\footnote{ziad@clermont.in2p3.fr}
\end{large} 

\bigskip
$^1$ 
Laboratoire de Physique Corpusculaire de Clermont-Ferrand, \\
IN2P3/CNRS Universit\'e Blaise Pascal, 
F-63177 Aubi\`ere Cedex, France\\

\noindent  
$^2$ 
I.N.F.N. - Sez. Genova,\\
Via Dodecaneso, 33, 16146 Genova, Italy \\  

\noindent

\vskip 1 true cm
\end{center} 
\vspace{1.0cm}  

\vspace{6pt}
\begin{center}{\large \bf Abstract}

We give a unitary description of T-odd effects in various sectors of atomic, nuclear 
and particle physics, like elastic scattering, deep inelastic processes and weak decays. 
This we get thanks to a particular transformation, which suggests a simple and 
unambiguous procedure for defining the two typical T-odd observables, the azimuthal 
asymmetry and the normal polarization. This approach allows to revise the two observables 
quite naturally in various situations, mainly in deep inelastic processes and in 
weak decays. In the latter case, useful suggestions for phenomenological analyses and 
an interesting result for normal and transverse polarization are derived.

\end{center}

\vspace{10pt}

\centerline{PACS numbers: PACS Nos: 11.30.Cp, 1.80.Et, 12.39Hg, 13.30.-a, 13.75.-n, 13.85.-t}
\newpage

\section{Introduction}
\label{intro}

 Time Reversal (TR) implies inversion of all spins and momenta involved in a process. 
 However, if performed without interchanging the initial and final state, this inversion 
 does not correspond to TR. It is sometimes called artificial TR\cite{si} and is connected 
 to important effects\cite{der}, called 
 T-odd\cite{bl}, produced by strong and electromagnetic interactions in two-body 
 scattering\cite{der}, deep inelastic processes\cite{br1,br2,si1,si2,co1,co2,bmp} and weak 
 non-leptonic\cite{va,wf,bl,bdl,arg} or semi-leptonic\cite{cgn,aos} decays. 
 Correspondingly, one has T-odd observables, typically polarizations and asymmetries, which 
 manifest themselves in all of the above mentioned physical situations, provided initial-state 
 interactions (ISI)\cite{si} or/and final-state interactions (FSI)\cite{si,wf} are involved. 
 In weak decays, the so-called fake\cite{ddu} T-odd terms are tightly entangled with real time 
 reversal violation (TRV)\cite{va}, which is of major interest to physicists, and it is not 
 a trivial problem to separate the two contributions\cite{va,wf,wf2}. 

 Such effects are largely studied in the literature\cite{ajd1}. Here we revisit the subject, 
 by means of a geometrical interpretation\cite{der}, which allows a unitary description of the 
 effects in different sectors of particle physics, but also in nuclear and atomic physics. In 
 particular, the geometrical interpretation suggests an unambiguous procedure for defining the  
 typical T-odd observables. In the case of weak decays, the method is successfully implemented by 
 the non-relativistic Zemach formalism\cite{ze}, which presents some advantages over the covariant 
 formulation in phenomenological analyses. 

 In section ~\ref{sec:3}  we show some examples of 
 T-odd amplitudes, both in elastic scattering and in weak decays. In section ~\ref{sec:4}  we introduce 
 a particular transformation, from which we derive in a natural way the T-odd azimuthal asymmetries 
 and polarizations. Section ~\ref{sec:5}  is dedicated to a detailed description of T-odd observables in 
 deep inelastic processes. Section ~\ref{sec:6}  is devoted to the same subject in weak decays, where 
 the effects of TRV and FSI are discussed. Lastly, in section ~\ref{sec:7}  we draw a short conclusion.
  
 \section{Examples of T-odd Amplitudes}
 \label{sec:3}

 Here we show some examples of processes involving T-even and T-odd amplitudes. The first one consists
 of elastic scattering, the others concern weak decays. 

 \subsection{Spin-Orbit Scattering}

 We consider, for the sake of simplicity, the case of elastic scattering of a beam of spin-1/2 particles 
 by a spin-0 fixed target in a given frame. The scattering amplitude in that frame is of the type
 \begin{equation}
 f(\theta) = \chi_f^{\dagger}{\hat f}(\theta) \chi_i, \label{sc1}
 \end{equation}
 where the $\chi$'s are the usual spinors, $\theta$ the scattering angle and
 \begin{equation}
 {\hat f}(\theta)  = f_0(\theta) + \vec{\sigma} \cdot \vec{\ell} f_1(\theta).
 \end{equation}
 Here $\vec{\ell}$ is the orbital angular momentum (oam) operator and $\vec{\sigma}$ $\equiv$ $(\sigma_1, 
 \sigma_2,\sigma_3)$, the $\sigma_i$'s being the Pauli matrices. By expanding
 $f_0(\theta)$ and $f_1(\theta)$  into Legendre polynomials, this operator results in\cite{ld}
 \begin{equation}
 {\hat f}(\theta) = f_e(\theta) + i \vec{\nu} \cdot \vec{\sigma} f_o(\theta). \label{scam}
 \end{equation}
 Here 
 \begin{eqnarray}
 f_e(\theta) &=& \frac{1}{2ip}\sum_{l=0}^{\infty}[(l+1)(e^{2i\delta_l^+}-1) + l (e^{2i\delta_l^-}-1)] P_l(cos\theta),
 \ ~~~~~ \ ~~~~~ \ ~~~~~ \ \label{tev} 
 \\
 f_o(\theta) &=& \frac{1}{2ip}\sum_{l=0}^{\infty}(e^{2i\delta_l^+} - e^{2i\delta_l^-}) P_l^1(cos\theta)
 \label{tdd} 
 \end{eqnarray}
 and
 \begin{equation}
 \vec{\nu} = \frac{\vec{p}_i \times \vec{p}_f}{|\vec{p}_i \times \vec{p}_f|},
 \end{equation}
 $\vec{p}_{i(f)}$ being the initial (final) momentum of the spin-1/2 particles, $|\vec{p}_i|$
 =  $|\vec{p}_f|$ = $p$. The phase-shifts $\delta_l^{\pm}$ refer to partial waves of angular momenta $j = l \pm 1/2$. 

  We note that the factor $i$ in front of the second term of eq. (\ref{scam}) is a consequence of the hermiticity of 
 the oam operator\cite{ajd1}; a different motivation was given in refs. \cite{va,dl}.

 A similar behavior is exhibited in the scattering of an electron by an ~ external ~ constant ~ electrostatic 
 ~ field\cite{ld1}. The scattering amplitude reads, in the Dirac formalism,
 \begin{equation}
 f = \frac{e}{4\pi} {\bar u}(p_f) \gamma_0 u(p_i) A_0,
 \end{equation}
 where  $A_0$ is the potential corresponding to the electrostatic field. 
 Using the standard representation for Dirac matrices, we get again eqs. (\ref{sc1}) and (\ref{scam}), with
 \begin{eqnarray}
 f_{e(o)}(\theta) &=&  \frac{e}{4\pi}  A_0  f'_{e(o)}(\theta),
 \\
 f'_e(\theta) &=& E+m_e + (E-m_e)cos \theta, 
 \\
 f'_o(\theta) &=& (E-m_e)sin \theta,
 \end{eqnarray}
 $m_e$ and $E$ being the mass and energy of the electron. This treatment can be  generalized in a 
 straightforward way to any 
 QED and QCD vertex. Indeed, the only T-odd product compatible with parity conservation has the form
 $\vec{S}\cdot \vec{p}_1\times\vec{p}_2$, where $\vec{S}$ and $\vec{p}_i$ ($i$ = 1,2) are, respectively, a 
 spin vector and two momenta involved in the process considered.

 \subsection{Weak Decays}
 
 Here we consider three weak decays. Obviously, the formalism developed - the non-relativistic Zemach\cite{ze} 
 formalism - may be extended to other weak decays where the initial and final particles have the same quantum 
 numbers as in the reactions considered. Other decays are analyzed in Appendix.

 \subsubsection{Decay $\Lambda_b \to \Lambda J/\psi$}

 In this case the decay amplitude in the rest frame of $\Lambda_b$ reads as 
 \begin{equation}
 A = \chi^{\dagger}_f [a \vec{\sigma}\cdot \vec{e}  +  b (\vec{p}\cdot \vec{e})^2
 + i c \vec{e} \times \vec{\sigma}\cdot \vec{p} + d (\vec{p}\cdot \vec{e})] \chi_i. \label{wk3}
 \end{equation}
 Here $\vec{e}$ is the polarization vector of the $J/\psi$, $\vec{p}$ the momentum of the $\Lambda$ in the
 $\Lambda_b$ rest frame, $\chi_i$ and $\chi_f$ are the Pauli spinors of the initial and final baryon 
 and $a$, $b$, $c$ and $d$ are complex  amplitudes. The partial waves involved are $l$ = 0 (first term), 1 
 (third and fourth term) and 2 (second term). The first (third) term of eq. (\ref{wk3}) is apparently T-even 
 (T-odd); however, as we shall see in a moment, if the parent resonance is polarized, each such term 
 yields both T-even and T-odd contributions.  
 
 \subsubsection{A three-body decay: $\Lambda_b \to \Lambda \pi^+ \pi^-$}

 The decay amplitude may be parametrized as
 \begin{equation}
 A = \chi_f^{\dagger} (a + b \vec{\sigma} \cdot \vec{p}_+ 
  + b' \vec{\sigma} \cdot \vec{p} + i c \vec{\sigma}\cdot \vec{p}_+\times \vec{p}) \chi_i. 
 \label{wk7}
 \end{equation}
 Here $\vec{p}$ is the momentum of $\Lambda$  in the $\Lambda_b$ rest frame and $\vec {p}_+$ the 
 momentum of $\pi^+$ in the ($\pi^+$, $\pi^-$) rest frame. Moreover the complex amplitudes $a$, $b$, 
 $ b'$ and $c$ depend in this case on non-negative powers of $\vec {p}^2$, of $\vec {p}_+^2$ and of
 $\vec {p}_+ \cdot \vec {p}$.

 \subsubsection{Decay $\Lambda_c \to \Lambda \pi^+$}
 
 The usual parametrization for the decay amplitude reads
 \begin{equation}
 A = \chi_f^{\dagger} (a +  b \vec{\sigma} \cdot \vec{p}) \chi_i, 
 \label{wk8}
 \end{equation}
 with notations analogous to those used in the two cases above. 
 Here it is not immediate to recognize a T-odd amplitude. However, if the parent resonance is polarized - 
 as usually occurs in the production of hyperons and heavy baryons - we may repeatedly exploit the identity
 \begin{equation}
 \chi_i = \vec{\sigma} \cdot {\hat s} \chi_i,  
  \label{wk9}
 \end{equation}
 where ${\hat s}$ is a unit vector in the direction of the polarization vector of the initial baryon. As a result we  
 get 
 \begin{equation}
 A = \chi_f^{\dagger} \{ a +  b [\hat{s} \cdot \vec{p} + \frac{i}{2}  \vec{\sigma} \cdot 
 \vec{p} \times \hat{s} - \frac{1}{2} \vec{\sigma} \cdot 
 (\vec{p} \times \hat{s}) \times \hat{s}]\} \chi_i. \label{wk10}
 \end{equation}
 Here, as we shall see in detail in the next section, the first and second term within square brackets correspond 
 respectively to the longitudinal and  
 normal component of the polarization; the third one is T-even and has a nontrivial transverse component
 if  $\hat{s}$ is neither parallel nor perpendicular to $ \vec{p}$. The procedure just adopted may be generalized 
 to any  weak decay of a polarized fermion, therefore the final spinning products present a normal and a 
 transverse polarization \cite{da}, related to each other by a given amplitude. We note also that eq. (\ref{wk10}) 
 is alternative to famous Lee-Yang's\cite{ly} formula for the proton polarization in the $\Lambda$ decay.  
 
 \subsubsection{A Remark}
 
 We note that, as in the case of elastic scattering, also in the weak decay amplitudes, the T-odd
 term contains the factor $i$, due to the hermitian character of the oam operator. In the case of the decay 
 $\Lambda_c \to \Lambda \pi^+$, the factor $i$  has been deduced as a consequence of the properties of the Pauli 
 matrices. In this connection, we observe that T-odd observables are generated by the interference between T-even
 and T-odd amplitudes. In a decay, these correspond respectively to waves of different parity, as can be seen in 
 the examples above; therefore it does not occur in strong or electromagnetic decays.
  
 \section{T-odd Asymmetries and Polarization}
 \label{sec:4}

 The  probability of a given process is the modulus square of the corresponding amplitude, say $A$.
 Set
 \begin{equation}
 A = A_e + A_o,  
 \end{equation}
 where $A_e$ and $A_o$ are respectively T-even and T-odd amplitudes. The corresponding probability,
 ${\cal P} = |A|^2$, may be split into a T-even and a T-odd part:
 \begin{equation}
 {\cal P} = {\cal P}_e+{\cal P}_o, \label{todob}
 \end{equation}
 where
 \begin{equation}
 {\cal P}_e = |A_e|^2 + |A_o|^2,  \ ~~~~~~ \ ~~~~~~ \ {\cal P}_o = 2 \Re(A_eA_o^*). \label{itf}
 \end{equation}
 An artificial TR\cite{si} - to be denoted as $\bar{T}$ from now on - leaves ${\cal P}_e$ 
 unchanged, while inverting the sign of ${\cal P}_o$. 
 
 Now we state an important property of T-even and T-odd observables involved in physical 
 situations of the type described in sect. ~\ref{sec:3}. According to the examples above, a T-odd 
 interaction is described by a triple product of the type
 \begin{equation}
 \vec{S}\cdot \vec{p}_1\times\vec{p}_2 \ ~~~~~~ \ {\mathrm or} \ ~~~~~~ \ 
 \vec{S}_1\cdot \vec{S}_2\times\vec{p}. \label{tpco}
 \end{equation} 
 Here $\vec{p}$ and $\vec{S}$ are respectively momenta and spins of the particles involved in the 
 process considered. We stress again that, if parity is conserved, only the 
 former product may appear. If the triple products are non-trivial, any pair of these vectors 
 singles out a physically significant plane. We fix one such plane, say
 $\Pi$, and perform a ${\bar T}$ inversion and a $\pi$-rotation, named $R_{\perp}$, of the 
 overall system around the direction normal to $\Pi$. The new process is such 
 that\footnote{Apart from an overall phase factor in the amplitude.}

 - the components of the spins and of the momenta on that plane are T-even and are the same as for 
 the initial process;

 - on the contrary, the components perpendicular to $\Pi$ are T-odd and are inverted.

 Therefore, if there is an interference between the T-even and the T-odd amplitude, the two 
 processes have different probabilities, the difference corresponding to a T-odd observable.
 If, among the vectors which appear in (\ref{tpco}), the one not
 lying in the plane $\Pi$ is a momentum  ${\vec p}$ of a given particle, one has an 
 {\bf azimuthal asymmetry} with respect to $\Pi$, of the type
 \begin{equation}
 A = \frac{N(\phi)-N(-\phi)}{N(\phi)+N(-\phi)} \propto sin\phi,
 \end{equation} 
 where $\phi$ is the azimuthal angle of ${\vec p}$ with respect to $\Pi$ and $N(\phi)$ is the number of 
 events with an azimuthal angle between $\phi$ and $\phi+d\phi$. If, instead, $\Pi$ excludes the spin of 
 a particle, one can detect a {\bf normal polarization} with respect to that plane. Our line of reasoning 
 generalizes the considerations of ref. \cite{der}. 
 
 For this purpose, it is worth stressing, and exploiting, some kind of analogy between $\bar{T}$ and $P$.
 Indeed, consider again the polarized baryon decays studied in subsection 2.2. Since parity is violated, both 
 the {\bf transverse} and the {\bf longitudinal} polarization of the final baryon - two T-even observables -
 may be interpreted geometrically, by performing the transformation $P R_{\perp}$, $R_{\perp}$ being
 performed with respect to the plane
 singled out by the momentum of one of the decay particles and by the polarization of the parent resonance. 
 This leaves the momenta of the decay particles and the normal polarization unchanged, while inverting 
 transverse and longitudinal ones. Therefore, since the hamiltonian of the system is not invariant under the 
 transformation, and since parity-even and parity-odd amplitudes interfere, a weak decay of a polarized fermion 
 to spinnning particles presents in general  a longitudinal, a normal and a transverse polarization in the final 
 products. Analogously, if one or both decay products are unstable, and if the momentum of one of the final 
 particles is determined, it can be shown that a  {\bf transverse} and a {\bf longitudinal} asymmetry - both T-even 
 - occur\cite{ajd2}.  

 \section{T-odd Observables in Elastic Scattering \\
          and in Deep Inelastic Processes}
  \label{sec:5}

\subsection{Elastic Scattering} 

 In the case considered in subsection 2.1, we have two significant planes, $\Pi_1$ and $\Pi_2$, respectively
 normal to $\vec{p}_i \times  \vec{S}_i$ - where $\vec{S}_i$ is the initial polarization of the electron - and to
$\vec{p}_i \times \vec{p}_f$. According to the considerations above, the final 
 electron is expected to exhibit, on the one hand, an azimuthal asymmetry in the distribution with 
 respect to $\Pi_1$ and, on the other hand, a polarization normal to the plane $\Pi_2$, even in the 
 absence of an initial polarization. Both T-odd quantities are proportional to the interference term
  \begin{equation}
 I = -2 Im (f_e^*  f_o) \vec{\pi}  \cdot \vec{\nu}, \ ~~~~~~ \   \vec{\pi} =
  \chi_i^{\dagger} \chi_f  \chi_f^{\dagger} \vec{\sigma} \chi_i, \label{intt}
 \end{equation} 
 as follows from eqs. (\ref{sc1}) and (\ref{scam}). Averaging and summing respectively over initial and 
 final polarizations, one sees that the T-odd part of the differential cross section - which gives rise to the
 azimuthal asymmetry - is non-zero, either if there is an initial polarization, or if a final polarization 
 $\vec{P}_f$ is detected\footnote{In the latter case, the significant plane is normal to $\vec{p}_f \times  
 \vec{P}_f$.}. This last reads as 
  \begin{equation}
  \vec{P}_f  \propto -2 Im (f_e^*  f_o) Tr (\chi_i \chi_i^{\dagger} \chi_f  \chi_f^{\dagger} \vec{\sigma} \cdot \vec{\nu}
  \vec{\sigma}) + .... , \label{npol}
 \end{equation}
 dots referring to the T-even terms;
 this confirms that the normal component of the polarization is present even in absence of an initial polarization. 
 However, expansions (\ref{tev}) and (\ref{tdd}) imply that, in order to have a sizable  asymmetry or polarization, 
 the phase-shifts must be sufficiently large. Indeed, at first order Born approximation, these observables 
 vanish, as can be seen from eqs. (\ref{intt}) and (\ref{npol}). The same result was obtained in ref. \cite{der}, 
 with different arguments. As we shall see, this is a general requirement for T-odd observables.
 
  \subsection{Semi-Inclusive Deep Inelastic Scattering}
 
 Analogous conclusions may be drawn for inclusive high-energy reactions of the type\cite{si,ds1,ds2,ds3}
 \begin{equation}
 \ell^- p \to \ell^- h X, \label{sid}
 \end{equation}
 where $\ell^-$ is an electron or a muon, $p$ a proton and $h$ a hadron. The basic scattering is
 \begin{equation}
 \ell^- q \to \ell^- q. \label{elm}
 \end{equation}
 Here $q$ is the active quark of the proton, which yields a jet and fragments into the hadron $h$.  
The following two effects occur:

 a) If the initial proton is transversely polarized, we have an azimuthal asymmetry (the Sivers
 effect\cite{si1,si2,si}) of the final quark and of the final lepton, with respect to the plane singled 
 out by the initial momentum of the lepton and the polarization vector of the proton. 
    
 b) The final quark and the final lepton have a polarization normal to the scattering plane (Boer-Mulders 
 effect\cite{bmp}), even with an unpolarized proton.

 The asymmetry and the polarization of the final quark are transmitted  to the hadron $h$, as we shall see 
 in subsection 4.4. 
 
 Similarly, the $e^+ e^-$ annihilation into two hadron jets and the hadron-hadron deep inelastic collisions,
  whose basic processes are respectively  $e^+ e^-$ $\to$ $q \bar{q}$ and $q \bar{q}$ $\to$ $q \bar{q}$, 
 originate quarks polarized normally to the plane singled out by the initial beams and by the momentum
  of one of the quarks.

 As in the case of elastic scattering, the effects just described can be observed only if ISI and/or FSI 
 between active and spectator partons occur, so as to entail 
 a sufficiently large phase difference between the T-even and the T-odd amplitude. ISI and FSI may 
 arise from coherent long-range spin-orbit effects, like confinement or chiral symmetry breaking; 
 alternatively, in the framework of perturbative QCD, they are connected to power corrections, in which 
 case they are important only if the active quark is heavy\cite{si}.

 \subsection{Drell-Yan}

 This kind of reactions may be obtained from (\ref{sid}) and (\ref{elm}) by line reversal:
 \begin{equation}
 p {\bar h} \to \ell^- \ell^+  X, \ ~~~~~~ \ ~~~~~~ \  q {\bar q}\to \ell^- \ell^+. \label{dy} 
 \end{equation}
 The same considerations as above can be done about the influence of ISI on the T-odd effects.
 Moreover analogous asymmetries and normal polarizations are predicted\footnote{We limit our 
 treatment to low and intermediate energies, where the $Z$ exchange is negligible.}:

 a') With a transversely polarized proton, we have an azimuthal asymmetry of each lepton with respect 
 to the plane singled out by the momentum and the polarization vector of the proton (Sivers effect
 for Drell-Yan\cite{si1,si2}). 

 b') The two leptons are polarized normally to the plane which contains their momenta. 

 However, as already shown by Collins\cite{co2} with regards to the Sivers effect, the asymmetry and 
 the normal polarization in Drell-Yan are of opposite sign than in SIDIS. This can be  easily understood 
 by recalling the triple products that characterize such effects. For SIDIS we have an interference term 
 \begin{equation}
 I  \propto  \vec{\pi}_q  \cdot \vec{p}_i \times \vec{p}_f
 \end{equation}
 where $\vec{p}_i$ and $\vec{p}_f$ are the initial and final momentum of the lepton and $ \vec{\pi}_q$ 
 is defined by the second eq.  (\ref{intt}). On the contrary, for DY  
 \begin{equation}
 I' \propto  \vec{\pi}_q  \cdot \vec{p}_+ \times \vec{p}_-, 
 \end{equation}
 $\vec{p}_{\pm}$ being the momenta of the final leptons. But line reversal of the DY basic scattering 
 with respect to SIDIS implies a change of sign in the triple product\cite{co2}.

 \subsection{Collins Effect}

 Also the process of parton fragmentation, {\it e. g.}, 
 \begin{equation}
 q \to h ~ X. \label{col}
 \end{equation}
 exhibits T-odd effects. For example, as we have seen, in SIDIS
 the quark is polarized normally to the plane passing through the momentum  $\vec{p}_j$ of 
 the jet and the direction of the initial beam. Therefore the final hadron $h$ presents an azimuthal 
 asymmmetry with respect to this plane. Moreover, if $h$ is spinning, it is polarized normally to the 
 plane singled out by $\vec{p}_j$ and the momentum of the hadron.
 
 \section{T-odd Effects in Weak Decays}
 \label{sec:6}

The effects described above are present also in the weak decays considered in section  ~\ref{sec:3}, provided either
 the parent resonance is polarized or at least two of the decay products are spinning.  
 
 For example, in the decay $\Lambda_b \to \Lambda_{(c)} V (P)$\cite{aj} - where $V (P)$ is a vector (pseudoscalar) 
 meson - a significant plane $\Pi$ is normal to the cross product $\vec{S}_{\Lambda_b} \times \vec{p}_{V(P)}$; 
 then we expect a polarization of $\Lambda_{(c)}$ normal to this plane, or an azimuthal asymmetry in the 
 distribution of the secondary decay products with respect to $\Pi$. Furthermore, if the meson is a vector one,
 we expect a normal polarization also for this particle. A similar situation 
 occurs in the decay $B \to V_1 V_2$, considering the plane singled out by $\vec{S}_{V_1}$ and $\vec{p}_{V_1}$;
 some details and considerations about this case are given in Appendix.
  
 In the three-body decay $\Lambda_b \to \Lambda \pi^+ \pi^-$, we predict an azimuthal asymmetry in the 
 distribution of $\pi^+$ or of $\pi^-$, with respect to the plane normal to $\vec{S}_{\Lambda_b} \times 
 \vec{p}_{\Lambda}$. Similarly, in the semi-leptonic decay $\Lambda_b \to \Lambda \ell^+ \ell^-$, a polarization 
 of the final $\Lambda$, normal to the plane singled out by the momenta $\vec{p_{\Lambda}}$ and $\vec{p_{\ell^+}}$ 
 (or $\vec{p_{\ell^-}}$) in the $\Lambda_b$ rest frame, can be detected\cite{cgn,aos,asi}. 

 The decay amplitudes written in subsection 2.2 allow to calculate the asymmetries and polarizations as 
 functions of the partial amplitudes. As an example, we consider again the decay $\Lambda_b \to \Lambda_c P$,
 for which the interference between the first and the third term of eq. (\ref{wk10}) yields 
 \begin{equation}
 I_o \propto \Im (ab^*); \label{np}
 \end{equation}
 this gives the size of the well-known\cite{ly} normal polarization of the final baryon.
 Moreover, it confirms that, in order to have a sizable T-odd observable, the relative phase of the T-odd 
 amplitude to the T-even one must be sufficiently large. In a weak decay, this phase is constituted by two different 
 terms, as we are going to show in subsection 5.2. 

 Also two parity-odd, T-even interference terms occur in that decay, corresponding to the transverse polarization 
 and to part of the longitudinal one, as we have seen in sects. ~\ref{sec:3} and ~\ref{sec:4}. These contributions 
 are proportional to     
 \begin{equation}
 I_e \propto \Re (ab^*). \label{pvl}
 \end{equation}
  
 \subsection{Final State Interactions}

 The FSI play an important role in producing T-odd observables in weak decays, analogously to ISI and/or FSI in
 deep inelastic processes. Again, they may come either from coherent long-range 
 effects, or from the regime of perturbative QCD; the latter contribution becomes important in decays of heavy 
 hadrons, like $B$, $\Lambda_b$, {\it etc.}, especially in non-factorizing terms\cite{bb,ks,bn,dl}. But in 
 ~ charmless ~ decays of ~ hadrons containing the beauty quark, only light quarks are involved in the final state; 
 therefore this effect is negligible, as in the case of deep inelastic processes and we expect that such terms 
 do not contribute appreciably to the T-odd effects, often unwanted in the search for authentic TRV. Our 
 conclusions are analogous to those drawn by other authors\cite{bdl1,chg} and are confirmed experimentally 
 by the suppression of certain decay modes\cite{dld}. 

 \subsection{Time Reversal Violation}
 
 TR is generally violated in weak decays, therefore the various amplitudes are endowed with phases which are 
 not entirely due to strong or electromagnetic interactions. In particular, if the phases that cause TRV differ 
 according to the partial amplitudes, they have a nontrivial effect on the observables. As an example, consider 
 once more a decay of the type $(1/2)^+ \to (1/2)^+ 0^-$, like the third one of subsection 2.2, and let $ \Phi$ 
 be the relative phase of the amplitude $a$ to $b$. This consists of two terms,
 \begin{equation}
 \Phi = \delta + \varphi, \label{rlph}
 \end{equation}
 where $\delta$ is connected to strong and electromagnetic FSI, while $\varphi$ is a consequence of TRV. Then, 
 inserting eq. (\ref{rlph}) into (\ref{np}) and (\ref{pvl}) yields 
 \begin{equation}
 I_e \propto |a| |b| cos(\delta + \varphi),  \ ~~~ \ ~~~ \  
 I_o \propto |a| |b| sin(\delta + \varphi),    \label{ieo}  
 \end{equation}
 whence
 \begin{equation}
 tan (\delta+\varphi) \propto \frac{I_o}{I_e}. \label{tgw}
 \end{equation}
 Then we conclude that it is impossible to measure a
 TRV phase\cite{wf2} by analyzing a single decay of this kind. To do that, one has to compare data of that 
 decay with those of the CP-conjugated one. Indeed, denoting by ${\bar{I}_o}$ and ${\bar{I}_e}$ the analogous 
 quantities to (\ref{np}) and (\ref{pvl}) relative to the CP-conjugated decay, we have      
 \begin{equation}
 tan (\delta+\bar{\varphi}) \propto \frac{\bar{I}_o}{\bar{I}_e}, \label{tgc}
 \end{equation}  
  $\bar{\varphi}$ being the new TRV phase. If $I_e$ differs from ${\bar{I}_e}$, or/and 
 $I_o$ differs from ${\bar{I}_o}$, CP is violated. But a non-zero difference of $\varphi-\bar{\varphi}$ implies 
 also TRV, independent of the CPT symmetry\cite{pku,gh}.

 Assumption of such a symmetry entails  $\bar{\varphi} = -\varphi$, $|\bar{a}| = |a|$ and $|\bar{b}| = |b|$,  
 which allows to determine the weak phase, thanks to eqs. (\ref{tgw}) and (\ref{tgc}); moreover
 eqs. (\ref{ieo}) imply, together with the analogous ones for the CP-conjugated decay,
 \begin{equation}
 \bar{I}_e - I_e  \propto |a| |b| sin\delta sin\varphi, \ ~~~ 
 I_o - \bar{I}_o  \propto cos\delta sin\varphi. \label{dieo}  
 \end{equation}
 Then, in looking for TRV or CP-violating effects, T-odd quantities appear more convenient 
 than T-even ones, in that they do not vanish for $\delta \to 0$\cite{bl}. However, as shown before, T-even
 observables may provide information which is complementary to T-odd ones\cite{daj}.

 In this connection, we note that leptonic decays, like $\mu^- \to \nu_{\mu} e^- \bar{\nu}_e$ and $\tau^- \to 
 \nu_{\tau} \mu^- \bar{\nu}_{\mu}$, allow to avoid the problem of FSI, so that, {\it e. g.}, the normal
 polarization of the final lepton is due entirely to TRV. On the other hand, in semi-leptonic decays, like
 $n \to p e \bar{\nu}_e$, FSI can be taken into account to high accuracy\cite{ntrv}, since they are purely
 electromagnetic.     

  \subsection{Phenomenology}

 The triple product asymmetries and the normal polarizations are often exploited in the search for new physics
 beyond the Standard Model (SM) in weak decays, especially of heavy hadrons\cite{cf,bdl,bdl1,arg,dl,ddl,cgn,aos,asi,aj}. 
 Indeed, these quantities, generally small according to the SM predictions, become quite sizable  on 
 the basis of other models\cite{bdl,dl}, which involve more amplitudes with different phases and 
 comparable moduli. As we have just seen, such observables seem to be more indicated than those based, 
 {\it e. g.}, on differences of partial decay widths, since they do not need strong relative phases between the 
 interfering amplitudes, even they are enhanced by small values of those phases.

 Predictions of triple product asymmetries are often made at the level of quark-quark interactions\cite{bl,ddl,bdl1}, 
 which entails a dilution in the experimental results of polarizations and asymmetries. This is especially true 
 for meson decays\cite{dl,ddl}. For this reason, baryon decays appear more suitable\cite{bl,bdl,bdl1,arg}, 
 particularly those of the $\Lambda_b$'s arising from fragmentation of polarized $b$-quarks in $Z \to b {\bar b}$ 
 decays\cite{arg,bdl1}. Incidentally, a baryon produced in a deep inelastic collision has a normal polarization, as a 
 consequence of the Collins effect, illustrated in subsection 4.4.
 
 Moreover, one is often faced with a cascade decay, like $\Lambda_b \to (p \pi^-)_{\Lambda} V$, for which a 
 correlation of the type 
 \begin{equation}
 \vec{p}\cdot \vec{p}_1\times\vec{p}_2 
 \end{equation}
 may be used. Here $\vec{p}$  is the momentum of the parent resonance in the laboratory frame, 
 $\vec{p}_1$ the momentum of one of the intermediate resonances in the rest frame of the parent one, and 
 $\vec{p}_2$ the momentum of one of the decay products in the rest frame of the intermediate resonance.
 This correlation is more suitable than (\ref{tpco}) from the experimental viewpoint. However, in order to have a
 non-zero asymmetry, both the parent resonance and the intermediate one must be spinning and must decay
 weakly, as already found in a previous article\cite{ajd2}. Indeed, the asymmetry would be washed out by parity
 conservation in the secondary decay, whereas it is preserved by parity violation, which causes an asymmetry
 in the distribution of the final decay products with respect to the plane normal to $\vec{p}_1$. But this generates 
 a dilution in the T-odd asymmetry, which is therefore expected to be smaller than those associated to correlations 
 (\ref{tpco}). Analogous conclusions can be drawn from the numerical results of ref. \cite{arg}.

 To summarize, triple products are a useful tool for investigating possible hints to new physics. To this
 end, the analyses of heavy baryon decays appear indicated from various viewpoints; indeed, such baryons 
 are generally polarized and keep memory of the heavy quark polarization, moreover their decays do not give rise to 
 large FSI phases. 

\section{Conclusions}
 \label{sec:7}
 
 A geometrical interpretation of T-odd effects allows to establish an unambiguous procedure for defining, in very 
 different sectors of physics, the two typical and physically significant T-odd observables. An essential role in 
 such effects is played by the relative phase between the T-even and the T-odd amplitudes. As regards scattering and 
 deep inelastic processes, at least at not too high energies, this phase is due uniquely to strong and/or 
 electromagnetic ISI or FSI; on the contrary, in weak decays, it derives contributions also from TRV. A few, already 
 known, results concerning T-odd observables are re-derived, sometimes in a simpler way, in our approach. In weak 
 decays, our geometrical considerations, implemented with the non-relativistic Zemach formalism, lead to establishing 
 an interesting relation between the normal and the transverse components of the polarization of final baryons.
 Moreover, from the phenomenological point of view, that formalism suggests in a natural way the variables which 
 are most adherent to the triple products in the fundamental quark-quark interactions.

 \vskip 0.40in

 \setcounter{equation}{0}
 \renewcommand\theequation{A. \arabic{equation}}

 \appendix{\large \bf Appendix}
  
 Here we consider two weak decays which involve T-even and T-odd amplitudes. We use the Zemach 
 formalism, as an alternative to the covariant one.
  
 \vskip 0.15in
 \subsection*{\bf Decay $B \to V_1 V_2$}
 \vskip 0.15in
 Here $V$ denotes a vector meson. The decay amplitude reads, in the rest frame of $B$, 
 \begin{equation}
 A = a \vec{e}_1 \cdot \vec{e}_2 + b \vec{e}_1 \cdot \vec{p} ~ \vec{e}_2 \cdot \vec{p}
 + i c ~  \vec{e}_1 \times\vec{e}_2 \cdot \vec{p}, \label{wk1}
 \end{equation}
 $\vec{e}_{1,2}$ being the polarization vectors of the two final vector mesons, $\vec{p}$ the momentum of the
 meson $V_1$ and $a$, $b$, $c$ three complex amplitudes. The three terms correspond, respectively, to the
 partial amplitudes with $l$ = 0, 2 and 1, this last being T-odd. It is interesting to compare our parametrization 
 (\ref{wk1}) with the covariant one, proposed by other authors\cite{va,dl,ddl}, {\it i. e.},
 \begin{equation}
 A' = a' e_1 \cdot e_2 + \frac{b'}{m_B^2} p \cdot e_1 p \cdot e_2 + 
 i p \frac{c'}{m_B^2} \epsilon_{\mu\nu\rho\sigma} p^{\mu}q^{\nu}e_1^{\rho}e_2^{\sigma}. \label{wk2}
 \end{equation}
 Here the $e$'s are polarization four-vectors, $p$ the four-momentum of $B$ and $q$ = $p_2 - p_1$,  $p_i$
 being the four-momentum of the $i$-th vector meson ($i = 1, ~ 2$); moreover $a'$, $b'$ and $c'$ are related to $a$, $b$
 and $c$ by multiplicative constants. The relationship between the second term in (\ref{wk1}) and the
 corresponding one in eq. (\ref{wk2}) may be exhibited by recalling that, in the rest frame of $B$, 
 \begin{equation}
 p \cdot e_i = m_B e_i^0, \ ~~~ \ ~~~ \ 0 = p_i \cdot e_i =  E_i e_i^0 \mp \vec{p}\cdot \vec{e}_i, 
 \end{equation}
 the $\mp$-sign holding for $i=1 ~ (2)$. Taking into account that $E_i = 1/2 m_B$, it follows 
 \begin{equation}
 \vec{p}\cdot \vec{e}_i = \pm 1/2 ~Êm_B e_i^0 = \pm 1/2 ~ p \cdot e_i. 
 \end{equation}
 
 In order to understand intuitively how  normal polarization comes out in the decay of a spin-0 particle
 to two vector mesons, we recall that the quarks absorb and emit gluons, giving rise to spin-orbit 
 interactions, and so acquiring a normal polarization, as described in subsection 2.1. This is true in particular 
 for the quarks of the two different vector mesons, which therefore are polarized orthogonally to their momenta 
 and transfer this polarization to the hadrons.
 
  \vskip 0.15in 
 \subsection*{\bf A cascade decay: $\Lambda_b \to \Lambda_c^+ \pi^- \to (\Lambda \pi^+)_{\Lambda_c} \pi^-$}
 \vskip 0.15in
 The amplitude of this process reads
 \begin{equation}
 A = \chi_f^{\dagger} [(a + b \vec{\sigma}\cdot \vec{p}) \sum_j (\chi_{\Lambda_c})_j
   (\chi^{\dagger}_{\Lambda_c})_j (a' + b' \vec{\sigma}\cdot \vec{p}')] \chi_i. \label{wk4}
 \end{equation}
 Here the index $j$ runs over the values $\pm1/2$ of the third component of the $\Lambda_c$ spin along 
 some quantization axis; moreover $\vec{p}$ and $\vec{p}'$ are respectively the momentum of the 
 $\Lambda_c$ resonance in the $\Lambda_b$ rest frame and the $\Lambda$ momentum in the $\Lambda_c$
 frame. But 
 \begin{equation} 
 \sum_j(\chi_{\Lambda_c})_j(\chi^{\dagger}_{\Lambda_c})_j = 1/2(1+\vec{\sigma}
 \cdot \vec{P}_{\Lambda_c}),  
 \end{equation}
 where $\vec{P}_{\Lambda_c}$ is the polarization of the intermediate
 resonance $\Lambda_c$. By recalling the properties of the Pauli matrices, we get
 \begin{equation}
 A = \frac{1}{2} \chi_f^{\dagger}{\hat M}\chi_i, \label{wk5}
 \end{equation}
 with 
 \begin{eqnarray}
 {\hat M} &=& {\cal A} + {\cal B} \vec{\sigma} \cdot \vec{p} + {\cal C}\vec{\sigma}\cdot \vec{p}' 
 + {\cal D}\vec{\sigma}\cdot \vec{P}_{\Lambda_c} + 
 {\cal E}( \vec{p}\cdot \vec{P}_{\Lambda_c} \vec{\sigma}\cdot \vec{p}' - \vec{p} \cdot\vec{p}' ~ 
 \vec{\sigma}\cdot \vec{P}_{\Lambda_c}+ \nonumber
 \\
 &+& \vec{p}' \cdot \vec{P}_{\Lambda_c} \vec{\sigma}\cdot \vec{p})+ i [{\cal B} \vec{\sigma} \cdot \vec{p} 
 \times \vec{P}_{\Lambda_c} + {\cal C}\vec{\sigma}\cdot \vec{P}_{\Lambda_c} \times\vec{p}' + \nonumber
 \\
 &+&
 {\cal E}(\vec{\sigma} \cdot \vec{p} \times \vec{p}' + \vec{p} \times \vec{P}_{\Lambda_c} \cdot \vec{p}')] 
 \ ~~~~~~ \ ~~~ \ ~~~~~~ \ \ ~~~~~~ \ ~~~ \ ~~~~~~ \ \ ~~~~~~ \ \ ~~~~~~ \   
 \label{wk6}
 \end{eqnarray}
 and
 \begin{equation}
 {\cal A} = {\cal D}+{\cal E} \vec{p} \cdot \vec{p}', \ ~~~~~ \ {\cal B} = ba',
 \ ~~~~~~ \ {\cal C} = ab',  \ ~~~~~~ \ {\cal D} =  aa', \ ~~~~ \ {\cal E} = bb'. 
 \end{equation}

\end{document}